\documentclass[12pt]{article}

\usepackage{graphics,cite,amssymb,epsfig,float,psfrag}
\usepackage{rotating}
\usepackage{amsmath,amssymb}
\usepackage{slashed}
\usepackage{xcolor} 
\usepackage{graphicx}
\newcommand{\lsim}{\raisebox{-0.13cm}{~\shortstack{$<$ \\[-0.07cm] $\sim$}}~} 
\newcommand{\gsim}{\raisebox{-0.13cm}{~\shortstack{$>$ \\[-0.07cm] $\sim$}}~} 
\newcommand{\beq}{\begin{eqnarray}} 
\newcommand{\eeq}{\end{eqnarray}} 
\newcommand{\tb}{\tan \beta}

\begin{document}

\vspace{.8cm}

\hfill  LPT-Orsay/2015-63, KCL-PH-TH/2015-34

 
\vspace*{1.4cm}

\begin{center}

{\large\bf Into the multi--TeV scale with a Higgs golden ratio} 



\vspace*{.7cm}

{\sc Abdelhak Djouadi$^1$, Jeremie Quevillon$^2$} and {\sc Roberto Vega-Morales$^1$}

\vspace*{.7cm}

\begin{small}

$^1$ Laboratoire de Physique Th\'eorique, Universit\'e Paris--Sud and CNRS, F--91405 Orsay, France

\vspace*{.1cm}

$^2$ Theoretical Particle Physics \& Cosmology, Department
 of Physics, King's College London, 

London, WC2R 2LS, United Kingdom

\vspace*{.1cm}

\end{small}

\end{center}

\vspace*{1cm}

\begin{abstract} 

With the upgrade of the LHC, the couplings of the observed Higgs particle to fermions 
and gauge bosons will be measured with a much higher experimental accuracy than current measurements, but will still be limited by an order 10\% theoretical uncertainty.~In this paper, we re-emphasize the fact that the ratio of Higgs signal rates into two photons and four leptons, $D_{\gamma \gamma}= \sigma(pp\to H \to \gamma \gamma)/\sigma(pp\to H \to ZZ^* \to 4\ell^\pm)$ can be made free of these ambiguities.~Its measurement would be limited only by the statistical and systematic errors, which can in principle be reduced to the percent level 
 at a high--luminosity LHC.~This decay ratio would then provide a powerful probe of new physics effects in addition to high precision electroweak observables or the muon ${\rm g\!-\!2}$.~As an example, we show that the Higgs couplings to top quarks and vector bosons can be constrained at the percent level and that new Higgs or supersymmetric particles that contribute to the H$\gamma\gamma$ loop can be probed up to masses in the multi--TeV range 
and possibly larger than those accessible directly. 

\end{abstract} 

\thispagestyle{empty}

\newpage
\setcounter{page}{2}

\subsection*{1. Introduction} 

The newly begun LHC run will allow for a more thorough probing of the electroweak symmetry breaking (EWSB) scale and search for physics beyond the Standard Model (SM).~Most beyond the SM scenarios that address the so--called hierarchy problem such as supersymmetry, extra space--time dimensions, or composite models, predict the existence of new particles with masses at the TeV scale or below \cite{NP}.~The search for these new particles can of course be done by directly producing them and with the increase in center of mass energy by almost a factor of two, the LHC will indeed probe higher mass scales than currently.~The search could also be done indirectly by performing more precise measurements of the observed Higgs boson couplings, as they are sensitive to the virtual effects of the new particles.~These precision measurements will be allowed by the large increase in statistics that will be due not only to the upgrade in energy \cite{LHCXS,Baglio} but also by the expected upgrade of the integrated luminosity.~This will be particularly the case at a high--luminosity LHC (HL--LHC) in which one expects to collect a few ab$^{-1}$ data \cite{HL-LHC}, some two orders of magnitude more Higgs events than collected so far. 

The indirect search for new physics effects will be particularly efficient in two very clean
channels, the $H\to \gamma \gamma$ and $H\to ZZ^* \to 4\ell$ (with $\ell=e,\mu$) decay modes.~In fact, already at the previous run with $\sqrt s=7$--8 TeV and $\approx 25$ fb$^{-1}$ data, the signal strengths for these two channels have been measured with an accuracy at 
the level of 15\% if only the statistical and systematic errors are taken into 
account~\cite{ATLAS-all,CMS-all,ATLAS-CMS}.~At a HL-LHC, one expects that the statistical error, which is presently of ${\cal O}(10\%)$, will drop by more than an order of magnitude.~The systematic errors could also be significantly reduced.~Hence, an experimental accuracy at the percent level could in principle be achieved for these two channels at a HL--LHC.~Unfortunately, this precision will be spoiled by the large theoretical uncertainties that affect the two channels, which are expected to be $\mathcal{O}(10)\%$ \cite{LHCXS,Baglio}.

It has been advocated in many instances, in particular in Refs.~\cite{ratio-old,ratio}, 
that the theoretical uncertainties can be eliminated by performing ratios of signal rates 
when considering the same Higgs production mode (choosing exactly the same kinematical configuration) but different final state decay channels.~This is particularly the case for the ratio $D_{\gamma \gamma}= \sigma(pp\to H \to \gamma \gamma)/\sigma(pp\to H \to ZZ^*\to 4\ell^\pm)$ which will simply be given by the partial decay width into the very clean $H\to \gamma\gamma$ and $H\to 4\ell^\pm$ modes that have negligible theoretical uncertainty.~In fact, even some systematic errors such as the one due to the luminosity measurement, will cancel out in the ratio.~Hence, only some experimental (mainly statistical) uncertainties will be left implying that the decay ratio $D_{\gamma \gamma}$ could be measured at the percent level at a HL-LHC, as also indicated in Ref.~\cite{HL-LHC}.~This opens up an interesting possibility as this accuracy will be comparable to the expected size of effects from TeV scale new particles that could alter Higgs couplings and, in particular, the $H\gamma\gamma$ 
 loop induced vertex \cite{HHG,Anatomy}.~This would make $D_{\gamma \gamma}$ a comparable probe of new physics as some high--precision electroweak observables such as the $W$ boson mass $M_W$ or the electroweak mixing angle $\sin^2\theta_W$ as well as the muon anomalous magnetic moment $(g-2)_\mu$ \cite{PDG}.  
 
We emphasize in this note that a 1\% measurement of $D_{\gamma \gamma}$ would allow 
to probe new physics scales above a TeV, in some cases higher than 
those accessible in direct searches for new particles.~To illustrate 
this point, we examine various specific cases including the Minimal Supersymmetric Standard 
Model (MSSM)~\cite{HHG,Anatomy}, anomalous effective Higgs couplings to SM particles, and composite Higgs models.~Before exploring these possibilities, we first discuss the signal strengths in the $H\to \gamma\gamma, ZZ^*$ search channels and their ratio $D_{\gamma\gamma}$ as well as the present and expected precision at the LHC. 

\subsection*{2. The $\gamma\gamma$ and $4\ell^\pm$ signal strengths and 
D$_{\gamma\gamma}$ decay ratio}

Among the Higgs signal strengths in the various search channels at the LHC, defined as
\beq
\mu_{XX} = \sigma(pp \to H \to XX)/ \sigma(pp \to H \to XX)|_{\rm SM}  \label{muXX} ,
\eeq 
two have been precisely measured by the ATLAS and CMS collaborations: $\mu_{\gamma\gamma}$ and $\mu_{ZZ}$, corresponding to the very clean $H\to \gamma\gamma$ and $H\to ZZ^*\to 4\ell^\pm$ (with $\ell=e,\mu)$ final states.~The latest results quoted by the two experiments~\cite{ATLAS-all,CMS-all} with the $\approx 25$ fb$^{-1}$ data collected at $\sqrt s=7\!+\!8$ TeV are shown in Table 1, including the statistical (first) and the systematic (second) errors.~The latter involve systematic experimental errors, including $\approx 2.5\%$ error due to the luminosity measurement, as well as the theoretical uncertainty due to scale variation (to account for missing higher orders) and the errors in the parametrization of the parton distribution functions (PDFs) and the measurement of the strong coupling constant ($\alpha_s$).~These errors are all then combined in quadrature (instead of linearly as advocated in~\cite{LHCXS,Baglio}).~The impact of the theoretical uncertainties alone as assumed by ATLAS and CMS is shown in parentheses. 

\begin{table}[!h]
\renewcommand{\arraystretch}{1.35}
\begin{center}
\begin{tabular}{|c|c|l|}\hline
channel    & ATLAS & CMS \\ \hline
$\mu_{\gamma\gamma}$ & $1.17 ~ ^{+0.23}_{-0.23} ~ ^{+0.16}_{-0.11} ~~ (^{+0.12}_{-0.08})$ 
                     & $1.14 ~ ^{+0.21}_{-0.21}  ~ ^{+0.16}_{-0.10} ~~ (^{+0.09}_{-0.05})$ \\ \hline
$\mu_{ZZ}$           & $1.46 ~ ^{+0.35}_{-0.31} ~ ^{+0.19}_{-0.13} ~~ (^{+0.18}_{-0.11})$ 
                     & $0.93 ~ ^{+0.26}_{-0.23} ~ ^{+0.13}_{-0.09}$ 
 \\ \hline
\end{tabular}
\end{center}
\vspace*{-0.5cm}
\caption{The inclusive signal strengths in the two search channels $H\to \gamma\gamma$ and $ZZ$ as measured by ATLAS \cite{ATLAS-all} and CMS \cite{CMS-all} with the statistical and systematic (theoretical and experimental) errors indicated.~The theoretical uncertainty alone is shown in parentheses.} \label{table:muXX}
\vspace*{-0.2cm}
\end{table}

As can be seen in Table 1, the largest source of uncertainty at present is of statistical
nature and amounts to $\approx 20\%$ for $\mu_{\gamma\gamma}$ and $\approx 30\%$ for 
$\mu_{ZZ}$.~This error will be drastically reduced at the next LHC run as the Higgs 
data sample in these channels will be significantly increased.~In the main Higgs 
production process $gg\to H$ which generates more than $85\%$ of the total Higgs 
cross section even before kinematical cuts are applied, the cross section will increase by
a factor $\simeq 2.5$ when moving from a center of mass energy of 8 TeV to 14 TeV.~Assuming 3000 fb$^{-1}$ at a HL-LHC, one would have $300$ times more events than what has been
collected so far.~This sample will allow the reduction of the statistical errors quoted 
in Table 1 by a factor $\approx 20$, leading to an expected precision of $\mathcal{O}$(1--2\%) for $\mu_{\gamma\gamma}$ and $\mu_{ZZ}$.~The systematic uncertainties listed above are dominated by the theoretical ones as the experimental errors are rather small in these two cleanly measured channels. 

Assuming small statistical and experimental systematic uncertainties, one would be left only with the theoretical uncertainties which are mainly due to 
the unknown higher QCD orders (introducing a dependence on energy scale), the PDFs, and the determination of $\alpha_s$.~These represent the bulk of the systematic uncertainty quoted in Table 1 and which the ATLAS and CMS collaboration assume to be of order $\pm 10\%$.~In fact, at the time the measurement was made, this uncertainty was larger as the 
scale and PDF+$\alpha_s$ uncertainties in $gg\to H$ were both about $\pm 7.5\%$ for a $M_H=125$ GeV Higgs at $\sqrt s=8$ TeV (they stay at the same level at $\sqrt s=14$ TeV).~According to Refs.~\cite{LHCXS,Baglio}, as they are of theoretical nature, these
two uncertainties should not be treated as Gaussian and should be added 
linearly.~The total theoretical uncertainty should then be $\approx \pm 
15\%$ and hence larger than in Table 1.~Very recently, however, the QCD corrections have been 
derived at N$^3$LO \cite{N3LO}.~While these new corrections increase the cross section only
slightly, they reduce the scale dependence to less than 3\%, making 
the total theory uncertainty indeed closer to $\approx 10\%$. 

Nevertheless, there is a third source of uncertainties that has not been considered so far,
the one due to the use of an effective field theory (EFT) approach 
to calculate some corrections beyond next-to-leading order in the $gg\!\to \!H$ mode.~For example, the EFT approach cannot be used for the subleading $b$--quark loop contribution that gives a $\approx 10\%$ contribution when it interferes with the $t$--loop and which is currently known only to NLO accuracy \cite{ggH-NLO}.~Depending on how conservative one is, this uncertainty is expected to be in the range 4\% \cite{LHCXS} to 7\% \cite{Baglio}, bringing us back to the $\approx 15\%$ total uncertainty quoted above\footnote{Note that the theoretical uncertainties in the vector boson fusion (VBF) channel  are much smaller for the inclusive cross section, $\lsim 5\%$ \cite{LHCXS}.~However, the rate is an order of magnitude smaller than for gluon fusion even before the VBF cuts are applied, increasing the statistical error to ${\cal O}(5\%)$.~In addition, it was recently shown that the uncertainties are larger for the cross section with the VBF cuts \cite{VBF-cuts}.}.  

Furthermore, uncertainties in Higgs decay branching ratios, which are $\mathcal{O}(5\%)$ for both the $H\to \gamma\gamma$ and $ZZ^*$ decays \cite{BRerrors}, should be taken into account.~In fact, this uncertainty is solely due to the QCD ambiguities that affect the 
$H\to b\bar b $ partial width (mainly the parametric uncertainties due to the inputs 
$b$--quark mass and $\alpha_s$ coupling), that represents $\approx 60\%$ of the total 
width of a 125 GeV Higgs boson.~The uncertainties in the partial decay widths in all
the non--hadronic decay channels such as $H\to \gamma \gamma, ZZ$ are in turn completely 
negligible.~Hence, even if the theoretical knowledge of the various components
that make the $\mu_{\gamma\gamma}$ and $\mu_{ZZ}$ signal strengths improve by the time of a HL-LHC, the theoretical uncertainties will stay of order 10\%, thus limiting the power to probe TeV scale physics.
  
Of course all these problems disappear at once if we consider the ratio of the 
$H\to \gamma\gamma$ and $H\to ZZ^*$ signal strengths for a given production channel,
\beq
D_{\gamma\gamma} \! = \! \frac {\sigma ( pp \to H \to \gamma\gamma)}{ 
\sigma ( pp \to H \to ZZ^*)} \! =\!   \frac {\sigma ( pp \to H)\times {\rm BR} 
(H \to \gamma\gamma)}{ \sigma( pp \to H) \times {\rm BR} (H \to ZZ^*)} \! =\!  
\frac{\Gamma( H \to \gamma\gamma)}{ \Gamma ( H \to ZZ^*)}  \propto \frac{c_{\gamma}^2} 
{c_{Z}^2} .
\label{ratio} 
\eeq 
If the same kinematical configuration for the Higgs production mechanism is adopted for both $H\to \gamma\gamma$ and $H\to ZZ^*$, then the fiducial production cross section cancels in the ratio.~One is then left not only with the ratio of the branching ratios but with the {\it ratio of the partial Higgs decay widths}, as the total Higgs width cancels out in the ratio. The two partial widths are affected by very small theoretical uncertainties, presently below 1\% \cite{BRerrors}. In fact, to first approximation and up to some kinematical factors and small radiative corrections that are known with sufficient precision (in particular when $M_H$ will be more precisely measured), $D_{\gamma\gamma}$ will be simply given by the ratio of the squared ``reduced" Higgs coupling to photons and massive gauge bosons\footnote{Here, we assume $c_W\!=\!c_Z$ as is the case in the SM and most of its  extensions.~In fact, the ratio  $c_W/c_Z$ measures the breaking of custodial symmetry and is related to the $\rho$ parameter, $\rho \equiv M_W^2/(\cos^2 \theta_W M^2_Z)= c_{W}^2/c_{Z}^2$ which is very close to unity \cite{PDG}.~One can combine the results for $H\to ZZ$ and $H\to WW$ to increase the statistical accuracy if the systematics in the latter mode could also be made very small.}, $D_{\gamma \gamma} \propto c_{\gamma}^2/c_{V}^2$ where $c_X \equiv g_{HXX}/g_{HXX}^{\rm SM}$.

Another interesting aspect of this ratio is that systematic uncertainties common to both channels, such as the one due to the luminosity, will also cancel out leaving only the statistical errors.~Combining $\Delta \mu_{\gamma\gamma}$ and $\Delta \mu_{ZZ}$ in quadrature gives a statistical uncertainty of $\Delta D_{\gamma\gamma} \approx 2\%$ at the HL--LHC. 
In fact, the recently released ATLAS+CMS combined analysis of the Higgs properties gives a statistical (equivalent to total) uncertainty of $\Delta_{\rm Run1}^{\rm comb.} D_{\gamma \gamma}= ^{+26\%}_{-22\%}$ \cite{ATLAS-CMS}  which would then lead to an expected accuracy of 
$\Delta_{\rm HL\!-\!LHC}^{\rm comb.} D_{\gamma \gamma} \approx ^{+1.5\%}_{-1.2\%}$. 
Such a high level level of accuracy in Higgs physics, $\approx \pm 1\%$,  was envisaged only 
in the case of observables to be measured in the ``cleaner" environment of  future $e^+e^-$ colliders\footnote{In the past, the possibility to turn linear $e^+e^-$ machines 
into high--energy $\gamma \gamma$ colliders was considered with the motivation 
that it would allow for a measurement of the $H\gamma\gamma$ coupling at the 
few percent level \cite{Anatomy}.}. 

With this level of expected precision, the ratio $D_{\gamma\gamma}$ may provide a high precision electroweak observable at the LHC.~And because it involves the loop induced $H \to \gamma \gamma$ channel in which many charged particles could contribute, the decay ratio would allow us to probe more deeply the TeV scale as will be discussed with some examples below.

\subsection*{3. Probing effective couplings and a composite Higgs}

The ratio $D_{\gamma\gamma}$ measures the magnitude of the $H\gamma \gamma$ loop vertex normalized to the $HVV$ coupling.~In the SM, the former is generated by contributions from the $W$ boson and the heavy top quark (neglecting smaller contributions such as from the $b$--quark) which interfere destructively.~In beyond the SM scenarios, any particle that is electrically charged and couples to the Higgs boson will contribute to the $H\gamma\gamma$ loop.~However, contrary to SM particles which leave their imprint in the loop even if they are very heavy (as their coupling to the Higgs is proportional to the mass), heavy new physics (NP) particles will generically decouple from the $H\gamma \gamma$ vertex as $M_H^2/M_{\rm NP}^2$.~Nevertheless, if their coupling to the Higgs is ${\cal O}(\alpha_W)$, contributions of order 1\% could be achieved for masses $M_{\rm NP}\! \gsim\! 1\;$TeV.

The new physics contributions to $D_{\gamma\gamma}$ can enter either via deviations in the $W$ and top couplings to the Higgs or through `contact' Higgs-photon interactions generated via higher dimensional operators once heavy particle have been integrated out or in composite Higgs scenarios.~The relevant terms in the effective Lagrangian can be written as ($v = 246$~GeV),
\beq\label{eq:efflag}
\mathcal{L} = \frac{H}{v}
\Big(
c_V (2 M_W^2 W_\mu^+ W^{-\mu} + M_Z^2 Z_\mu Z^\mu)
- m_t \bar{t} (c_t + i \tilde{c}_t \gamma^5) t 
+ \frac{c_{\gamma\gamma}}{4}F^{\mu\nu}F_{\mu\nu}
+ \frac{\tilde{c}_{\gamma\gamma}}{4}\tilde{F}^{\mu\nu}F_{\mu\nu}
\Big) 
\eeq
where we have allowed for both CP even ($c_X$) and CP odd ($\tilde{c}_X$) couplings and assumed custodial symmetry to set $c_Z = c_W = c_V$.~At \emph{tree level} in the SM we have $c_V = c_t = 1$ and $c_{\gamma\gamma} = \tilde{c}_{t} = \tilde{c}_{\gamma\gamma} = 0$ while at
one--loop a contribution $c_{\gamma\gamma} \approx -0.008$ is generated.

The loop $W,t$ loop contributions and the effective $H\gamma\gamma$ interaction enter into $D_{\gamma\gamma}$ as,
\beq \label{eq:Dgaga}
D_{\gamma\gamma} \equiv 
\mu_{\gamma\gamma}/\mu_{ZZ} =
 \left|1.28\, - 0.28\, (c_{t}/c_V) + (c_{\gamma\gamma}/c_V) \right|^2 +
\left| 0.43\, (\tilde{c}_{t}/c_V) + (\tilde{c}_{\gamma\gamma}/c_V) \right|^2 \, 
\eeq
where the numerical values correspond to the $W$ and top loop functions~\cite{HHG,Anatomy} for $M_H=125$ GeV.~As is clear, $D_{\gamma\gamma}$ is only sensitive to the ratios of couplings $c_X/c_V, \tilde{c}_X/c_V$ and not the absolute magnitude of the couplings.~Note also that $D_{\gamma\gamma}$ can not lift degeneracies such as when $c_X \to -c_X$ and is not directly sensitive to CP violation which requires interference between the CP even and CP odd couplings\footnote{To be sensitive to such effects requires combination with other channels or the study of differential distributions as in $H\to 4\ell$, $2\ell\gamma$ or $H\to\gamma\gamma$ where the photons convert in the detector; see Refs.~\cite{Roberto}.}.~Below we examine new physics possibilities which can enter into $D_{\gamma\gamma}$ via $W$ and top couplings or effective $H\gamma\gamma$ couplings.

\begin{figure}[!h]
\vspace*{-.2mm}
\begin{center}
\includegraphics[scale=.41]{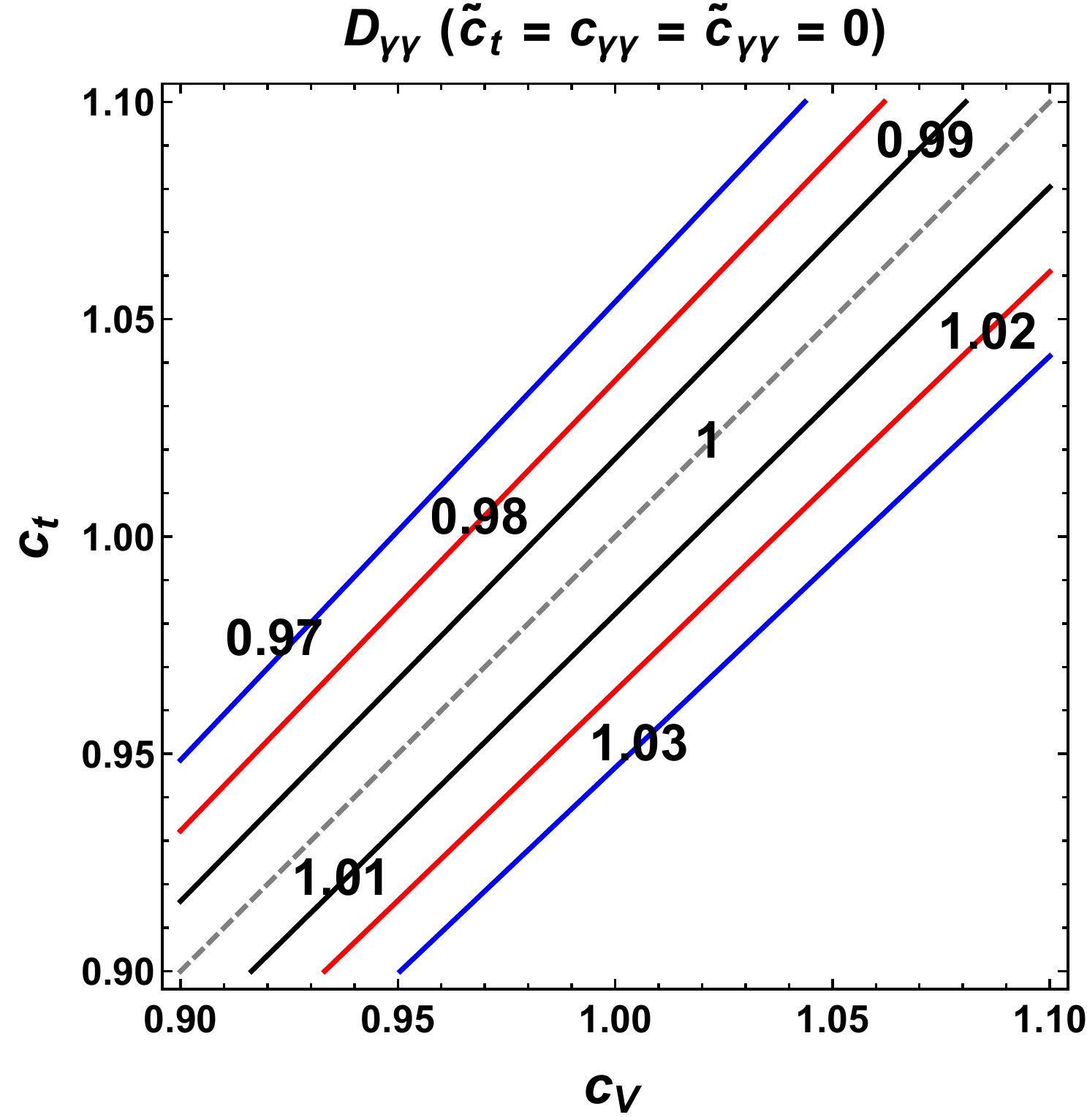} \hspace*{1cm} 
\includegraphics[scale=.36]{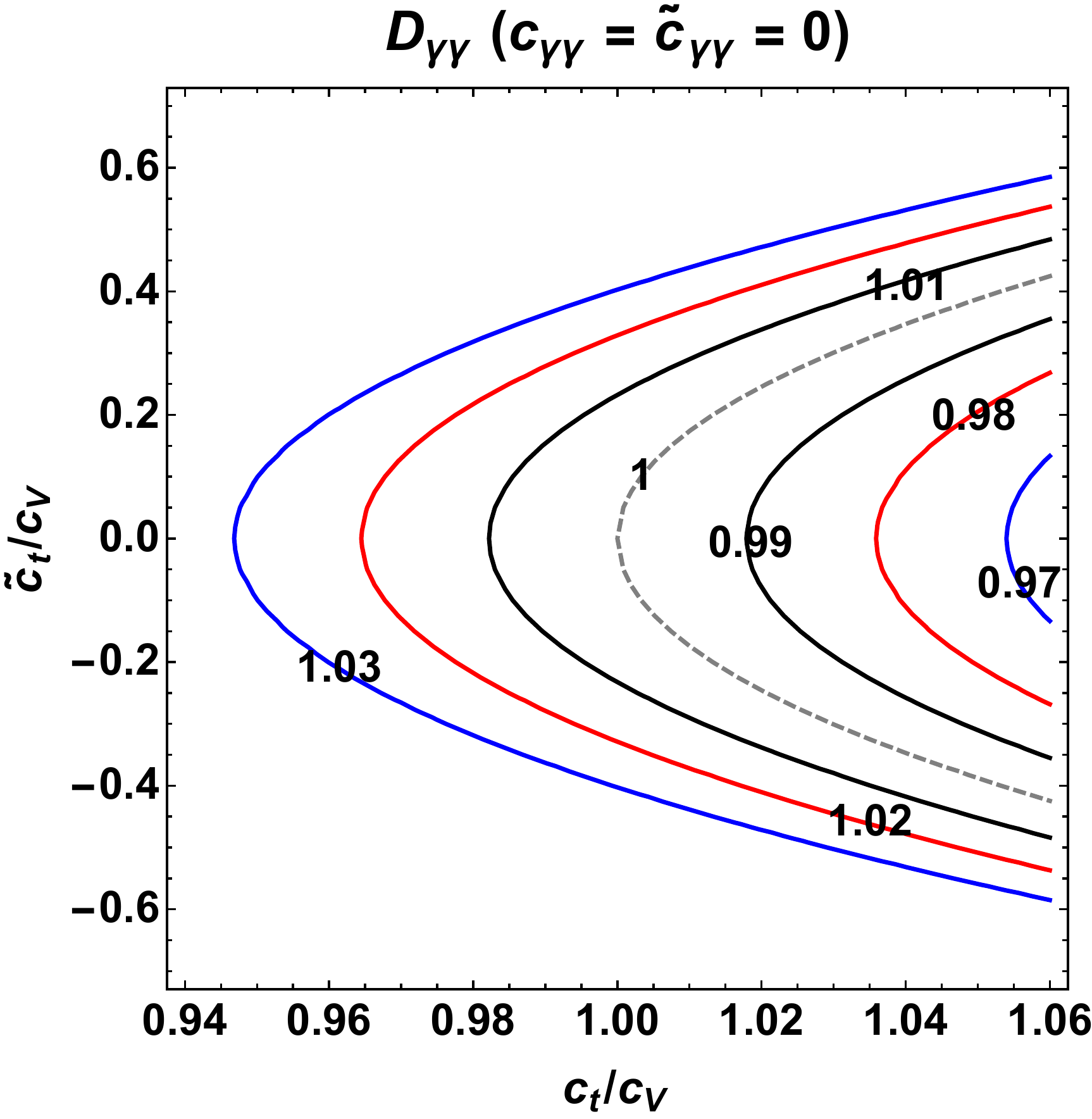}
\end{center}
\vspace*{-7mm}
\caption{Contours of $D_{\gamma\gamma} \! + \! \Delta D_{\gamma\gamma}$ in the $[c_V, c_t]$ (left) and $[c_{t}/c_V, \tilde{c}_{t}/c_V]$ (right) planes assuming $D_{\gamma\gamma} \!=\! 1$ and $\Delta D_{\gamma\gamma} = 1\%\, {\rm (black)},~2\%\, {\rm (red)},~3\%\, {\rm (blue)}$.}
\label{fig:ctVcv}
\vspace*{-1mm}
\end{figure}

Focusing first on deviations which enter through the top and $W$ couplings, we take $c_{\gamma\gamma} \!= \!\tilde{c}_{\gamma\gamma} \! = \! 0$ and study how well the $c_V$ and $c_t, \tilde{c}_t$ couplings can be constrained with $D_{\gamma\gamma}$.~We show in~Fig.~\ref{fig:ctVcv} contours of $D_{\gamma\gamma} \pm \Delta D_{\gamma\gamma}$ in the $[c_V, c_t]$ (left) and $[c_t/c_V, \tilde{c}_t/c_V]$ (right) planes assuming $D_{\gamma\gamma} = 1$ and $\Delta D_{\gamma\gamma} = 1\%, 2\%$ and $3\%$.~In the $[c_V, c_t]$ plane, focusing on the region around the SM point $c_t\! =\! c_V \!=\!1$ which is preferred by present data \cite{PDG}, one sees that $D_{\gamma\gamma}$ constrains $c_t$ and $c_V$ to lie in a narrow band.~Assuming $c_V =1$, $D_{\gamma\gamma}$ could constrain the CP--even top coupling at the level of $c_t \! \approx \;  $1--2\% for $\Delta D_{\gamma\gamma} \sim 1\%$.~Clearly for the $[c_t/c_V, \tilde{c}_t/c_V]$ plane, $D_{\gamma\gamma}$ alone is not enough to lift the degene
 racy which exists when $\tilde{c}_{t} \to -\tilde{c}_{t}$ even for $c_V = 1$.~However, if one assumes that the CP--odd coupling $\tilde{c}_t \approx 0$ 
(as suggested by measurements of electric dipole moments for instance \cite{McKeen}), 
then $D_{\gamma\gamma}$ would constrain the CP--even coupling ratio $c_{t}/c_V$ to  
$\sim 2\%$ for $\Delta D_{\gamma\gamma} \sim 1\%$.

\begin{figure}[!h]
\vspace*{-.2mm}
\begin{center}
\includegraphics[scale=.44]{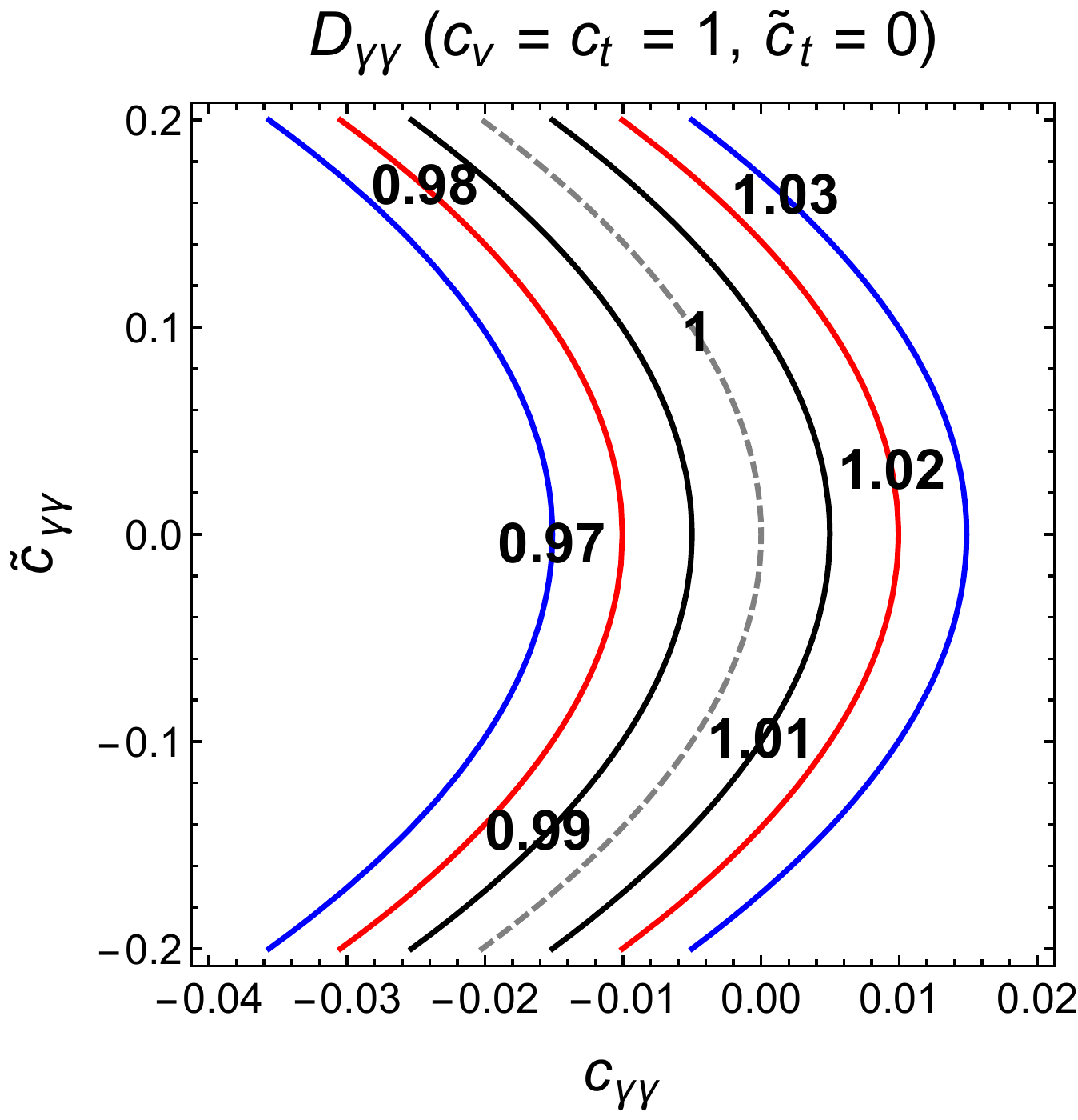}\hspace*{.4cm} 
\includegraphics[scale=.37]{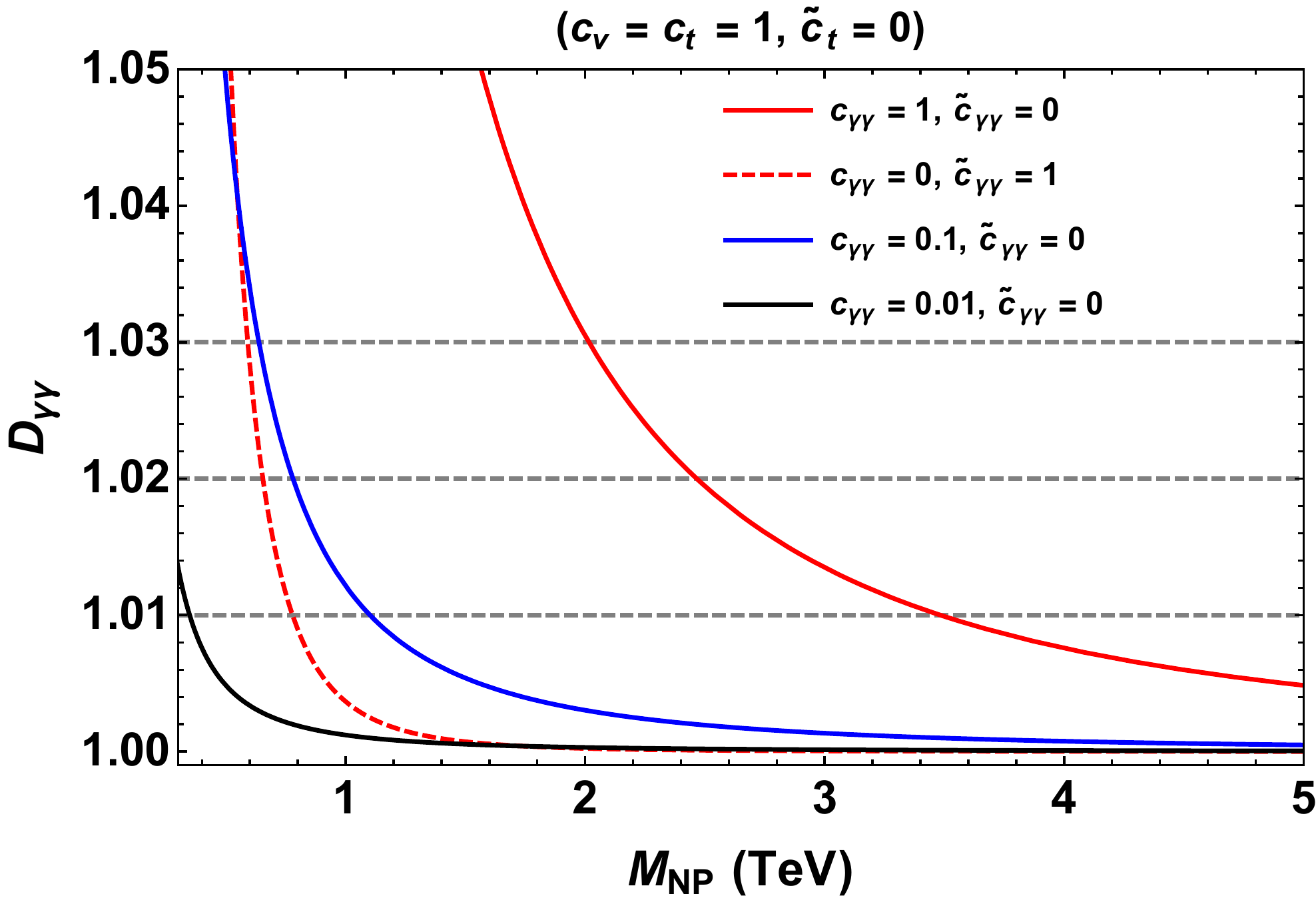}
\end{center}
\vspace*{-.7cm}
\caption{Left: As in Fig.~\ref{fig:ctVcv}, but in the $[c_{\gamma\gamma}, \tilde{c}_{\gamma\gamma}]$ plane.~Right: $D_{\gamma\gamma}$~vs.~$M_{\rm NP}$ for 
$c_{\gamma\gamma} \! = \! 1, \tilde{c}_{\gamma\gamma} \! = \! 0$ (red), 
$c_{\gamma\gamma} \! = \! 0, \tilde{c}_{\gamma\gamma} \! = \! 1$ (red-dashed), 
$c_{\gamma\gamma} \! = \! 0.1, \tilde{c}_{\gamma\gamma} \! = \! 0$ (blue) and 
$c_{\gamma\gamma} \! = \! 0.01, \tilde{c}_{\gamma\gamma} \! = \! 0$ (black)
and where we have rescaled the couplings as $c_{\gamma\gamma} \to c_{\gamma\gamma}v^2/M_{\rm NP}^2$ (see text).}
\label{fig:EFT}
\vspace*{-.2cm}
\end{figure}

Next, we examine the case where new physics enters only through the $H\gamma\gamma$ effective couplings while the $W$ and top couplings have SM--like values $c_V \! = \! c_t \! =\! 1$.~On the left hand side of of~Fig.~\ref{fig:EFT}, we display contours of $D_{\gamma\gamma} \!+\! \Delta D_{\gamma\gamma}$ in the $[c_{\gamma\gamma}, \tilde{c}_{\gamma\gamma}]$ plane again assuming $D_{\gamma\gamma} \!=\! 1$ and $\Delta D_{\gamma\gamma} = 1\%, 2\%, 3\%$.~Here also, $D_{\gamma\gamma}$ alone cannot lift all degeneracies but for $\tilde{c}_{\gamma\gamma} \approx 0$ \cite{McKeen}, one can potentially obtain percent level constraints on the ratio $c_{\gamma\gamma}/c_V$.

The effective couplings $c_{\gamma\gamma}, \tilde{c}_{\gamma\gamma}$ are implicitly associated to some new physics scale $M_{\rm NP}$ and scale as $v^2/M_{\rm NP}^2$, which suppresses the dimension--6 operators in the unbroken ${\rm SU(2)_L \! \times \! U(1)_Y}$ invariant phase from which they arise.~To visualize which scales can be probed, we perform the rescaling $c_{\gamma\gamma} \to c_{\gamma\gamma}v^2/M_{NP}^2$ and plot $D_{\gamma\gamma}$~vs.~$M_{\rm NP}$ for various values of $c_{\gamma\gamma}$ and $\tilde{c}_{\gamma \gamma}$ in the right of~Fig.~\ref{fig:EFT}.~We see for example that $\Delta D_{\gamma\gamma} \approx 1\%$ would potentially allow us to probe scales as high as $\sim 3$--4 TeV for $c_{\gamma\gamma} = \mathcal{O}(1)$.~For electroweak size couplings of $\mathcal{O}(10^{-2})$, only scales that are an order of magnitude smaller, $\sim 400$--500~GeV, can be probed.~Since it does not interfere with the dominant $W$--loop contribution, the sensitivity of the CP odd  
$\tilde{c}_{\gamma\gamma}$ coupling is less strong and even for $\mathcal{O}(1)$ couplings, only scales of $\lsim 1$~TeV can be probed with $\Delta D_{\gamma\gamma} \approx 1\%$.

\begin{figure}[!h]
\vspace*{-2mm}
\begin{center}
\includegraphics[scale=.45]{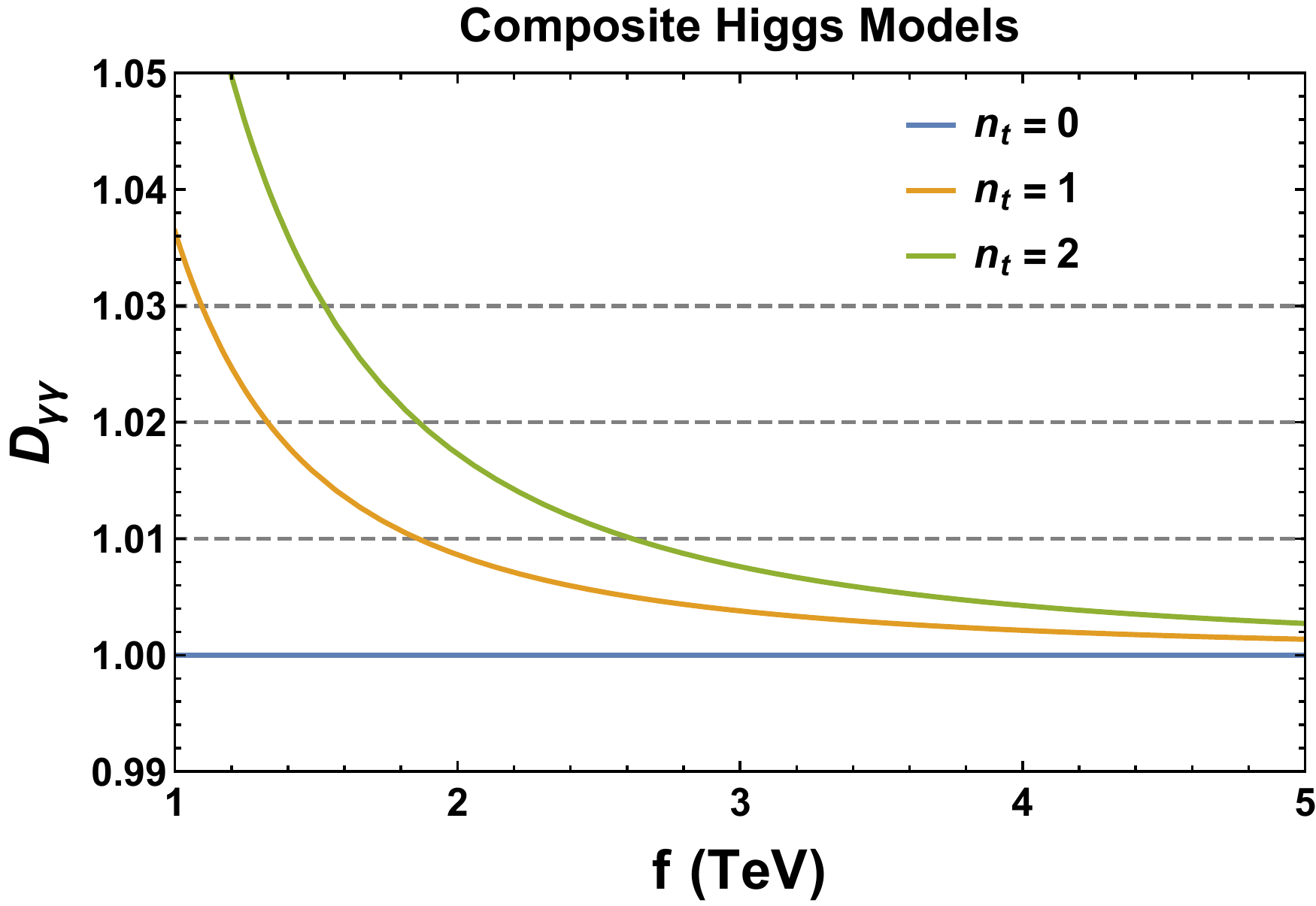}\hspace*{1cm} 
\includegraphics[scale=.32]{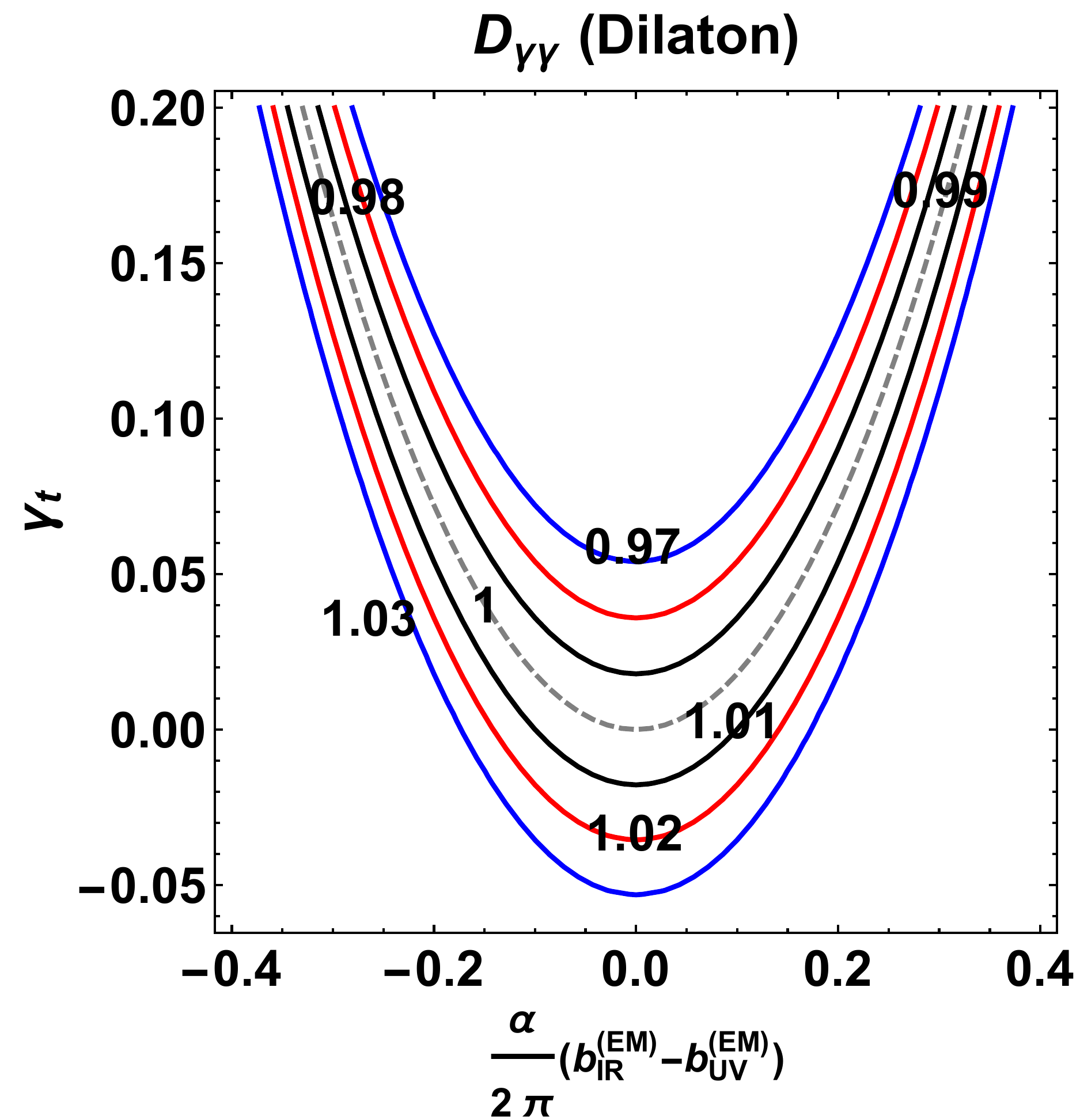}
\end{center}
\vspace*{-7mm}
\caption{
Left: $D_{\gamma\gamma}$ vs.~compositeness scale $f$ for $n\! =\! 0$ (blue), 
$n \!= \!1$ (orange) and $n\! =\! 2$ (green) in minimal composite Higgs scenarios.~Right:~contours of $D_{\gamma\gamma} \!+\! \Delta D_{\gamma\gamma}$ in the $[\alpha
/4\pi(b^{\rm (EM)}_{\rm IR} \!-\! b^{\rm (EM)}_{\rm UV}), \gamma_t]$ plane with $D_{\gamma\gamma} \!=\! 1$ and $\Delta D_{\gamma\gamma} \!= \!1\%$ (black), 2\% (red), 3\% (blue).}
\label{fig:dilaton}
\vspace*{-1mm}
\end{figure}

Finally, let us briefly discuss contributions to $D_{\gamma\gamma}$
in two generic composite Higgs scenarios.~The first case is when the Higgs boson arises as a pseudo-Goldstone boson of a spontaneously broken approximate global symmetry in a strongly interacting sector~\cite{MCHM}.~In this class of models, the $HVV$ couplings are given by $c_V = \sqrt{1 - \xi}$ where $\xi = v^2/f^2$ with $f$ the compositeness scale or the decay constant of the pseudo-Goldstone Higgs boson.~The couplings to fermions depend not only on the scale $f$, but also on the representations of the SM fermions under the global symmetry group of the strongly interacting sector for which there are many possibilities~\cite{comphiggs}.~For the minimal case based on the SO(5) global symmetry broken to SO(4) the fermion couplings take the form~\cite{comphiggs},
\beq
{c_t}/{c_V} = [1 - (1 + n)\xi]/({(1 - \xi)}),~~\tilde{c}_t = c_{\gamma\gamma} = \tilde{c}_{\gamma\gamma} = 0 ,
\eeq
where $n$ is a positive integer dictated by the fermion representation.~In~Fig.~\ref{fig:dilaton} (left), we display $D_{\gamma\gamma}$ as a function of the compositeness scale $f$ for $n = 0, 1, 2$.~One first notes that $D_{\gamma\gamma}$ is not sensitive to the $n = 0$ scenario since in this case the dependence on $\xi$ drops out.~For the cases of $n =1$ and $n = 2$, we see that with a $1\%$ measurement of $D_{\gamma\gamma}$, compositeness scales of 2--3~TeV can potentially be probed.

As a second scenario, we consider the case of a dilaton where the Higgs state arises as a pseudo-Goldstone boson of spontaneously broken scale invariance~\cite{dilaton}.~The dilaton again couples to gauge bosons and fermions but it also has a direct coupling to photons which is generated by the trace anomaly.~Since all these couplings depend on $\xi$ in the same way, $\propto \sqrt{\xi}$, the ratio of widths $D_{\gamma\gamma}$ is not directly sensitive to the composite scale $f$.~The coupling ratios entering $D_{\gamma\gamma}$ are explicitly given by~\cite{dilaton},
\beq
{c_t}/{c_V} = (1 + \gamma_t),~~
{c_{\gamma\gamma}}/{c_V} = {\alpha}/{(4\pi)} \times (b^{\rm EM}_{\rm IR} - b^{\rm EM}_{\rm
UV}),~~\tilde{c}_t = \tilde{c}_{\gamma\gamma} = 0 ,
\eeq
where $\gamma_t$ is the anomalous dimensions which measures the explicit breaking due to the mixing in the ultraviolet (UV) between the elementary and composite states associated with the top quark.~The $\beta$ function coefficients, $b^{\rm EM}_{\rm UV}$ and $b^{\rm EM}_{\rm IR}$, parameterize the explicit breaking of scale invariance in the ultraviolet and the infrared (IR) due to the contribution of composite fields to the running of the photon gauge coupling~\cite{dilaton}.

In~the right--hand side of Fig.~\ref{fig:dilaton}, we show contours of $D_{\gamma\gamma} + \Delta D_{\gamma\gamma}$ in the $[\alpha/4\pi(b^{\rm (EM)}_{\rm IR} - b^{\rm (EM)}_{\rm UV}), \gamma_t]$ plane assuming again $D_{\gamma\gamma} = 1$ and the usual $\Delta D_{\gamma\gamma}$ precision of 1\%, 2\% and 3\%.~One can see that for $b^{\rm (EM)}_{\rm IR} - b^{\rm (EM)}_{\rm UV} \approx 0$, as in the SM~\cite{dilaton}, $\gamma_t$ can be constrained at the 1--2\% level for $\Delta D_{\gamma\gamma} = 1\%$.~As we approach this level of precision it will start to become possible to exclude negative values of $\gamma_t$ for any $b^{\rm (EM)}_{\rm IR} - b^{\rm (EM)}_{\rm UV}$.~There is also a degenerate second region at $\gamma_t \sim 7$ which is not displayed.

\subsection*{4. Probing the heavy new particles of the MSSM}
 
We now examine one--loop contributions to the Higgs decay to photons from new heavy charged particles in the MSSM.~As is well known, the MSSM possesses a two Higgs 
doublet structure that leads to a physical spectrum with five Higgs states: 
two CP--even $h$ and $H$ (with $h$ being the observed one), a CP--odd $A$ and two charged states $H^\pm$~\cite{HHG,Anatomy}.~Two parameters are needed to describe this sector: $M_A$ and $\tan\beta$, the ratio of vacuum expectation values of the two Higgs fields.~This is true not only at tree--level but also when higher order corrections are included, provided that the constraint $M_h = 125$ GeV is used as in the $h$MSSM approach discussed in Ref.~\cite{hMSSM}.~The couplings of $h$ to fermions and gauge bosons, when 
normalized to the SM Higgs couplings are simply given by,
\begin{eqnarray}  
c_V  \! = \!  \sin(\beta \!-\! \alpha)  \ ,  \ \  c_t \!  = \!  
\cos  \alpha/ \sin\beta \ , \ \   c_b \!  = \!  -\!
\sin  \alpha/ \cos\beta ,
\label{cplg}
\end{eqnarray}
with $\alpha$ the mixing angle in the CP--even Higgs sector which, in the $h$MSSM, is simply given in terms of $M_A,\tb$, and $M_h$.~When $M_A \! \gg \! M_Z$, one is in the decoupling regime in which $\alpha \approx \beta- \frac{\pi}{2}$ and $h$ has SM--like couplings, $c_t\!=\!c_b\!=\!c_V\!=\!1$.~In this decoupled regime, which is also implied by the experimental data~\cite{PDG}, the heavier charged Higgs states have mass $M_{H^\pm} \!= \!\sqrt { M_A^2+M_W^2}$ 
while $H$ and $A$ have comparable masses and couplings. 

We first consider MSSM contributions to the $h\gamma\gamma$ loop
induced vertex\footnote{In the limit where the loop particle is much heavier than the $h$ state, the contributions to the $h\to \gamma \gamma$ vertex up to coupling and charge factors, are $A_1 = -7$ for spin--1 states and $A_{1/2}= +\frac43$ for spin--$\frac12$ fermions; the contribution of a spin--0 particle in the same configuration would be $A_{0}= -\frac13$ \cite{HHG,Anatomy}.~Note that while spin--1 and spin-0 states decouple like $1/M_{\rm NP}^2$, spin--$\frac12$ fermions decouple only as $1/M_{\rm NP}$.} \cite{gamma} 
in the limit where all superpartners are very heavy.~In this case two effects are at play.~First, the charged $H^\pm$ state will contribute to the $h\gamma 
\gamma$ amplitude, $ \mathcal{M}_{H^\pm} \propto -\frac13 g_{hH^+ H^-} M_W^2/M_{H^\pm}^2$.~A second contribution also occurs at low $M_{H^\pm}$ values as one is outside the decoupling regime and the reduced Higgs couplings $c_W$ and $c_t$ are not SM--like, 
eq.~(\ref{cplg}).  

\begin{figure}[!h]
\vspace*{-4mm}
\begin{center}
\includegraphics[scale=0.4]{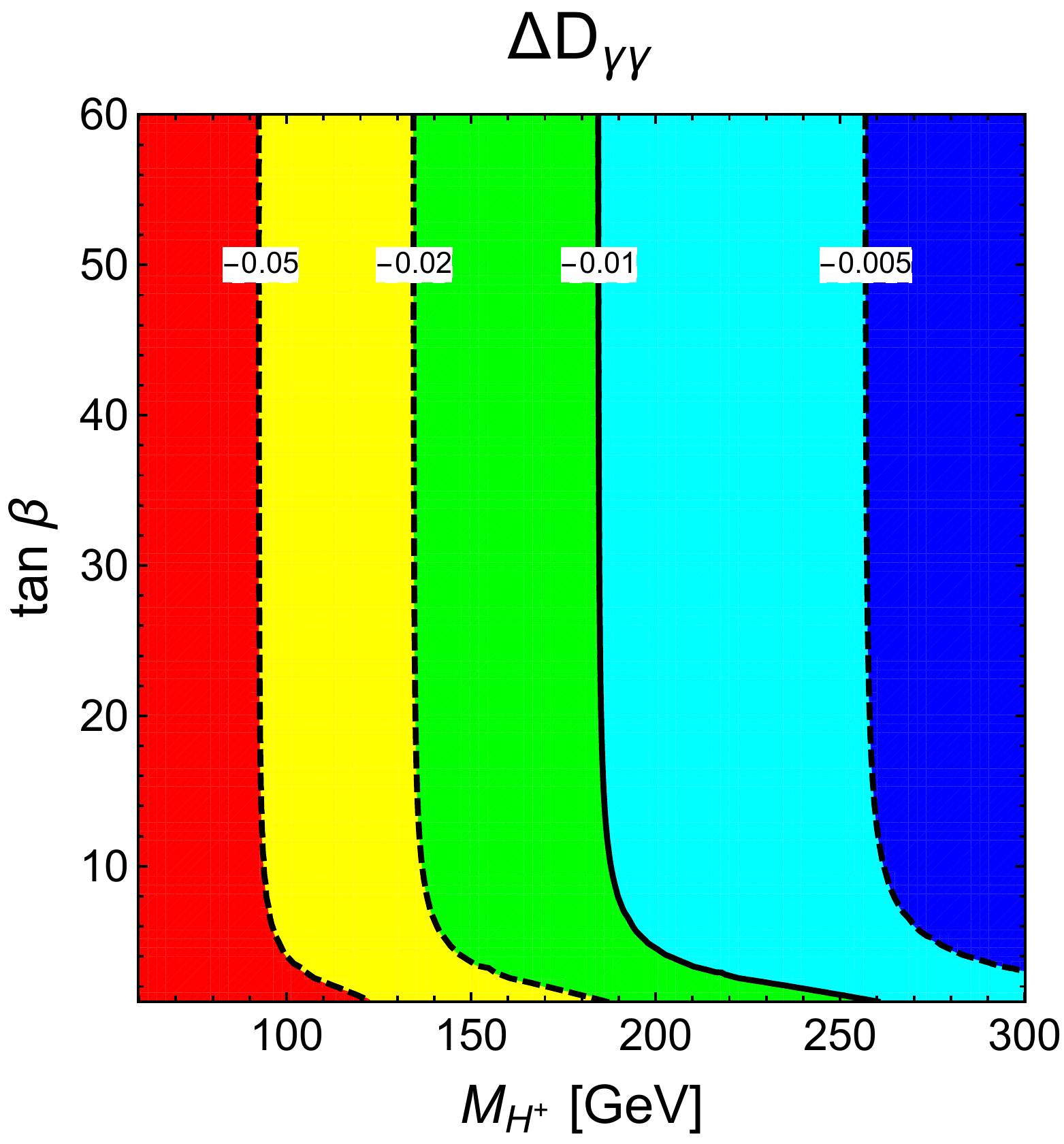} \hspace*{5mm}
\includegraphics[scale=0.4]{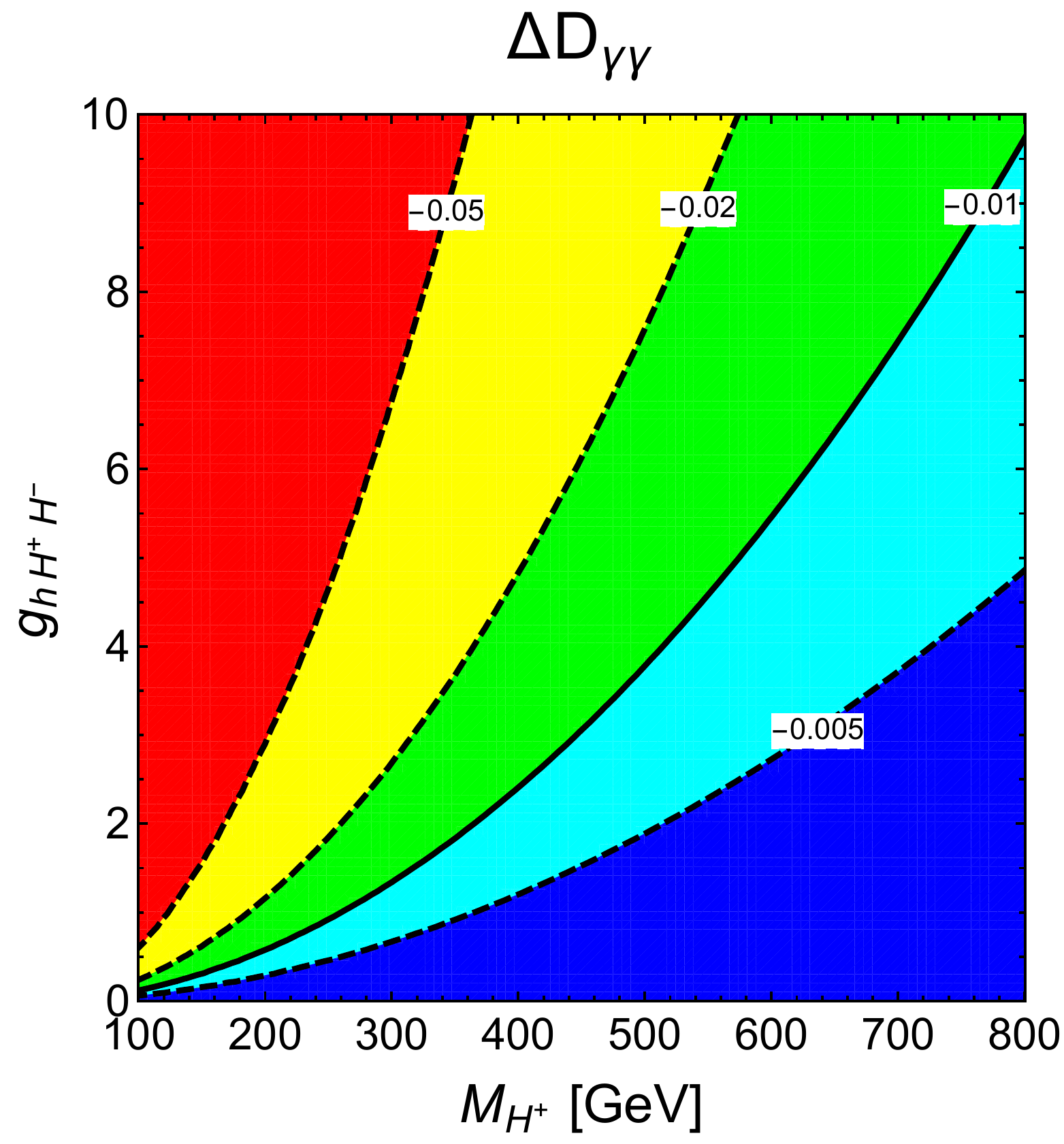} 
\end{center}
\vspace*{-.8cm}
\caption{Left:~Contributions of the MSSM Higgs sector to $D_{\gamma\gamma}$ in the
$[M_{H^\pm}, \tb$] plane when all superparticles are assumed to be heavy.~Right:~Contributions to $D_{\gamma\gamma}$ from charged Higgs bosons in a 2HDM with arbitrary ${hH^+ H^-}$ coupling in the $[M_{H^\pm}, g_{hH^+ H^-}]$ plane when the ``alignment" limit $\sin(\beta-\alpha)=1$ is considered.}
\label{fig:H+}
\vspace*{-.3cm}
\end{figure}

The simultaneous impact of these two contributions is illustrated in the 
$[M_{H^\pm}, \tb$] plane in Fig.~\ref{fig:H+} (left).~As can be seen, a 1\% measurement of $D_{\gamma\gamma}$ probes only $H^\pm$ masses of ${\cal O}(200$ GeV), especially at high $\tb$.~This is due to the fact that the decoupling limit $c_V=c_t=1$ is already reached for such a $H^\pm$ mass and the fact that the $h H^+ H^-$ coupling is small in the MSSM.~Indeed, at tree--level, it is given by ($c_W^2= 1- \sin^2\theta_W \approx 3/4$),
\beq
g_{h H^+H^+} = \sin(\beta-\alpha) + \cos2 \beta \sin(\beta+\alpha) /(2c_W^2) 
\stackrel{M_A \gg M_Z} \to 1 - \cos^2 2 \beta/(2c_W^2) ,
\eeq
which approaches $\approx 1/3$ for $\tb \gg 1$.~The $D_{\gamma\gamma}$ sensitivity to the
$H^\pm$ states is only slightly improved for values of $\tb\approx 1$ when the coupling 
becomes of order unity.~In contrast, in a general two--Higgs doublet model (2HDM) \cite{charged}, the coupling $g_{h H^+ H^-}$ is essentially a free parameter and can be larger,
leading to more significant contributions to $D_{\gamma\gamma}$ \cite{HHG,charged}.~This is illustrated in the right-hand side of Fig.~\ref{fig:H+} where the $H^\pm$ contribution is 
displayed in the $[M_{H^\pm}, g_{hH^+ H^-}]$ plane for a 2HDM in the ``alignment" limit 
$\sin(\beta-\alpha)=1$ which leads to $c_V=c_t=1$.~In this case $H^\pm$ masses close
to the TeV scale can be probed for $g_{hH^+ H^-} \gsim 5$ if $\Delta D_{\gamma\gamma}
\approx 1\%$.~Note that in triplet Higgs models where doubly charged $H^{\pm\pm}$ states
are present, the contributions to $D_{\gamma\gamma}$ can be even larger~\cite{H++}.~In this case, values as large as $M_{H^{++}} \approx 2.5$ TeV can be probed for $g_{hH^{++} H^{--} } \approx 5$ if an accuracy of $\Delta D_{\gamma\gamma}\! \approx\! 1\%$ is achieved. 

Turning to the effects of the superpartners, we assume for simplicity that we are in the decoupling regime $M_A\! \gg\! M_Z$ with $h$ having SM--like couplings, and focus on the direct contributions to the $h\gamma\gamma$ loop.~As the superpartner couplings 
to the $h$ state are not proportional to their masses, the loop contributions are damped 
by powers of $M_{\rm NP}$.~Three contributions can be important in the MSSM (besides that from $H^\pm$ discussed earlier) \cite{gamma}:~those from the charginos \cite{inos}, the tau slepton \cite{Tau}, and the stop squark, see e.g.~\cite{Stop}.~We discuss each of them in the phenomenological MSSM \cite{pMSSM} in which all soft SUSY--breaking parameters are free but with the constraint that they are not CP or flavor violating.~The program HDECAY \cite{hdecay}has been used for the numerical analysis. 

\begin{figure}[!h]
\vspace*{-2mm}
\begin{center}
\begin{tabular}{cc}
\hspace*{-1.5cm}
\includegraphics[scale=0.48]{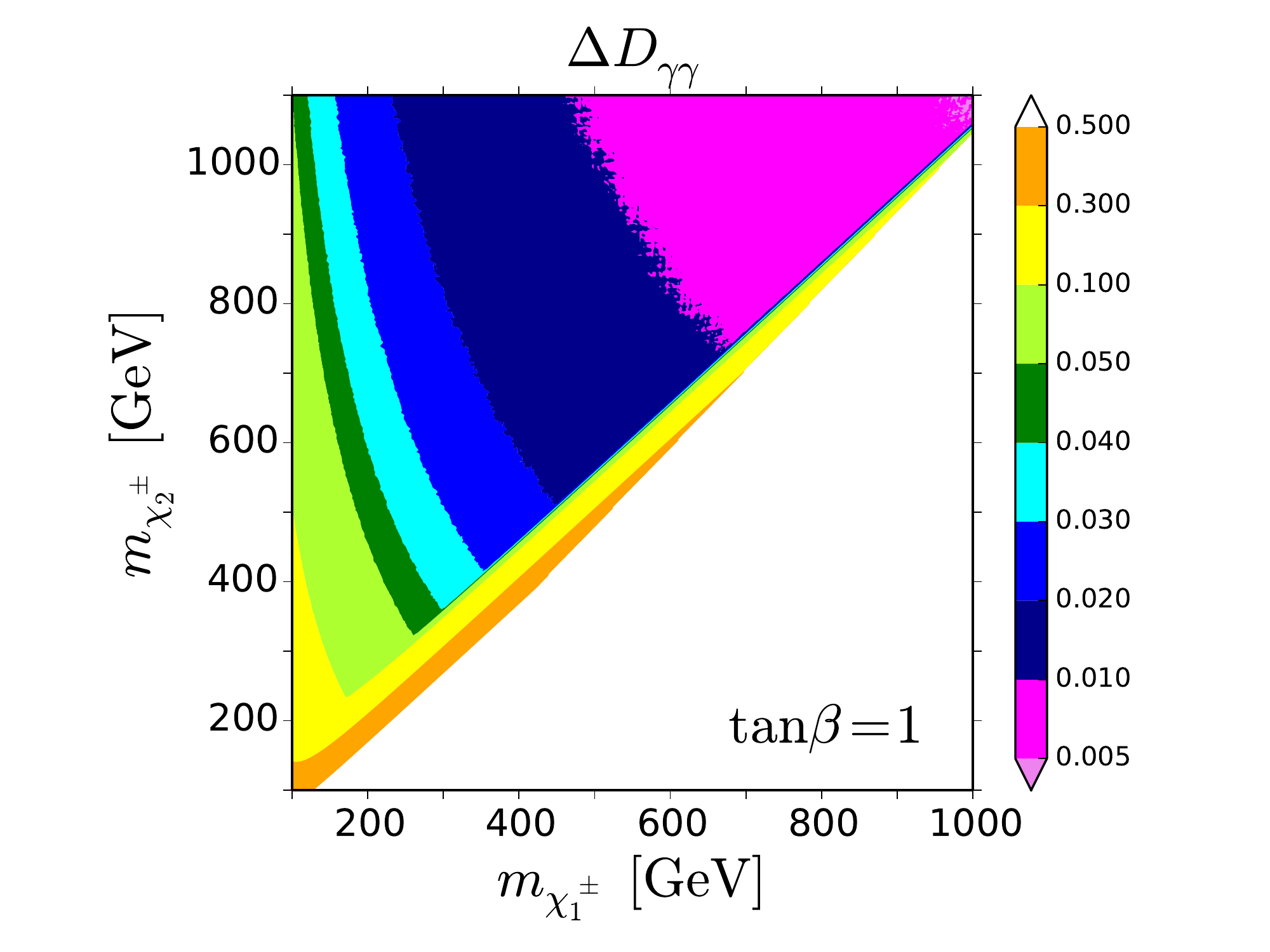}~~~~~ &
\hspace*{-1.5cm}
\includegraphics[scale=0.43]{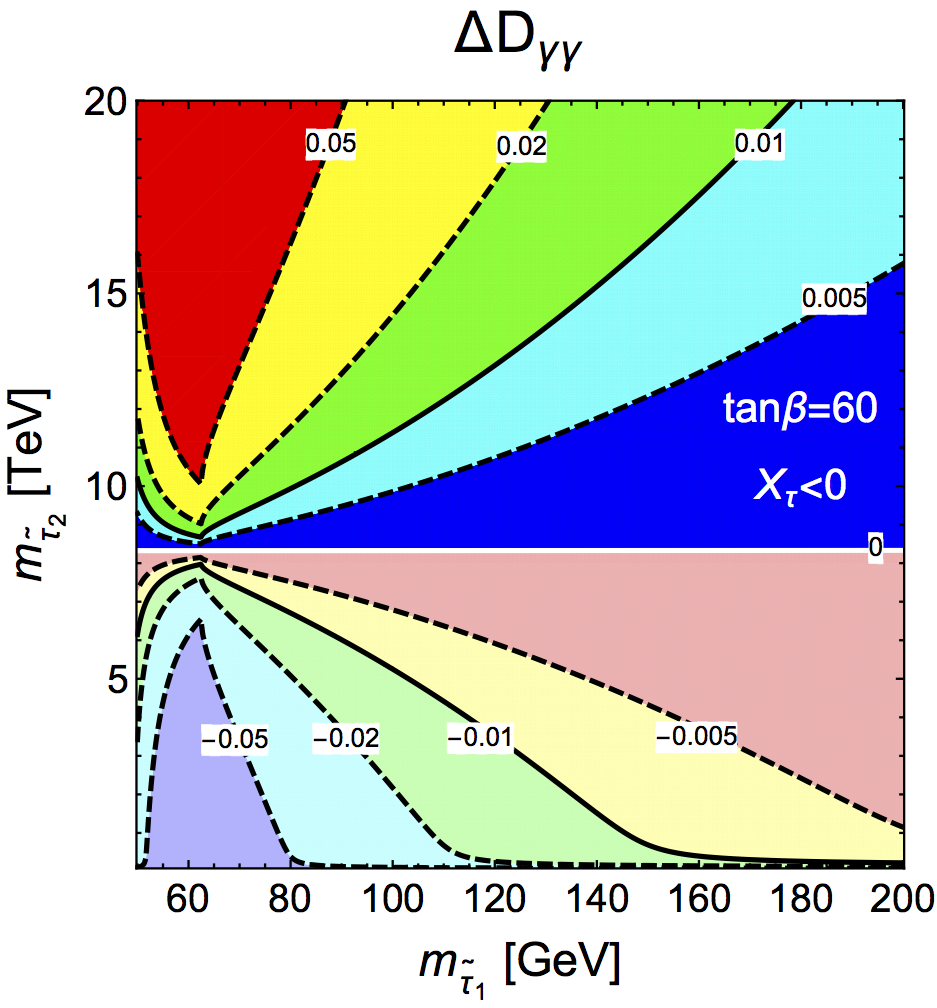} 
\end{tabular}
\vspace*{-.5cm}
\caption{Contours for the contributions of the charginos (left) and stau leptons 
(right) to $D_{\gamma\gamma}$ in the planes formed by the masses of two states.~For charginos
$\tb=1$ is assumed while for staus, we set $\tb=60 , X_\tau \leq 0$ and assume $m_{\tilde \tau_L}=
m_{\tilde \tau_R}$.}
\label{fig:inos}
\end{center}
\vspace*{-.8cm}
\end{figure}

$i)$~Light charginos: the chargino system is described by the gaugino and higgsino 
mass parameters $M_2$ and $\mu$ and by $\tb$.~If $M_2\gg |\mu|$, the lightest chargino 
$\chi_{1}^\pm$ is a pure higgsino while the heavier one $\chi_{1}^\pm$ a pure gaugino;
for $M_2\ll  |\mu|$ the situation is reversed.~The contributions of the spin--$\frac12$ 
charginos to the $h\gamma\gamma$ vertex scales like $\propto g_{h \chi_i^+ \chi_i^-}/m_{
\chi_i^\pm}$ with the $h \chi_i^+ \chi_i^-$ couplings being of electroweak strength 
and maximal when $\tb \approx 1$ and the $\chi_i^\pm$ states are equal mixtures of 
higgsinos and gauginos.~Their contribution to $D_{\gamma\gamma}$ is shown in the 
left--hand side of Fig.~\ref{fig:inos} in the [$m_{\chi_1^\pm},  m_{\chi_2^\pm}]$
plane for fixed $\tb =1$ (the sensitivity of $D_{\gamma\gamma}$ is lower
at higher $\tb$ values).~One sees that for $\Delta D_{\gamma\gamma} \approx 1\%$, 
chargino masses of the order of $m_{\chi_1^\pm} \approx 0.5$ TeV and 
$m_{\chi_2^\pm} \approx 1 $ TeV can be probed.

$ii)$ Stau sleptons (the contribution of first/second generation sleptons is negligible \cite{gamma}): the stau system can be described with three parameters, the soft SUSY--breaking mass parameters $m_{\tilde \tau_L}$ and $m_{\tilde \tau_R}$, and the mixing parameter $X_\tau= A_\tau-\mu \tb$.~In the decoupling limit and assuming $m_{\tilde \tau_L}=m_{\tilde \tau_R}$,
the Higgs--stau coupling reads $g_{h\tilde \tau \tilde \tau}\! = \! -\frac{1}{4}\cos2\beta+\frac{m_{\tau}^2}{M_{Z}^2}+\frac{m_{\tau}X_{\tau}}{2M_{Z}^2}$. 
When $X_\tau$ is large and negative, the coupling simplifies to $g_{h\tilde \tau \tilde \tau}\! \propto \! m_\tau X_\tau$ and is important only for large $X_\tau$, making the splitting between the two  staus also very large.~This allows for one of them to be rather light rendering maximal the impact of the $\tilde \tau_1$ loop in the $h\gamma\gamma$ vertex, which interferes constructively with the $W$ loop.~When the mixing parameter is not negative enough, the Higgs--stau coupling is then positive and its contribution interferes destructively with the dominant one coming from the $W$ boson.~Nevertheless, as the contribution of a spin--0 particle is small and damped by $g_{h\tilde \tau \tilde \tau}/m_{\tilde \tau}^2$, the  staus decouple quickly from the amplitude.~This is exemplified in Fig.~\ref{fig:inos} (right) where the contribution to $D_{\gamma\gamma}$, assuming $m_{\tilde \tau_L}=m_{\tilde \tau_R}$ is displayed in the plane $[m_{\tilde \tau_1}, m_{\tilde \tau_2}]$.~We see that  staus of a few hundred GeV could still contribute by more than 1\% to $D_{\gamma\gamma}$.~One should note that  staus are almost undetectable at the LHC in direct searches \cite{Tau}. 

$iii)$ Stop loops:~they provide the largest contribution to the $h\to \gamma\gamma$ vertex.
The stop sector, similarly to the stau sector, can be parametrised by the three inputs
$m_{\tilde t_L}, m_{\tilde t_R}$ and $X_t= A_t-\mu/\tb$ (the SUSY scale is defined
as $M_S= \sqrt {m_{\tilde t_L} m_{\tilde t_R}}$ and should be of order 1 TeV for
a mixing $X_t/M_S \approx 2$ in order to allow for an $h$ boson mass of $M_h\!=\!125$ GeV \cite{Arbey}).~If the mixing parameter is large, the two stop masses will split and $\tilde t_1$ will be much lighter than $\tilde t_2$ with large coupling to the $h$ state, $g_{h \tilde t_1 \tilde t_1} \propto m_t X_t$.

\begin{figure}[!h]
\vspace*{-4mm}
\begin{center}
\begin{tabular}{cc}
\includegraphics[scale=0.5]{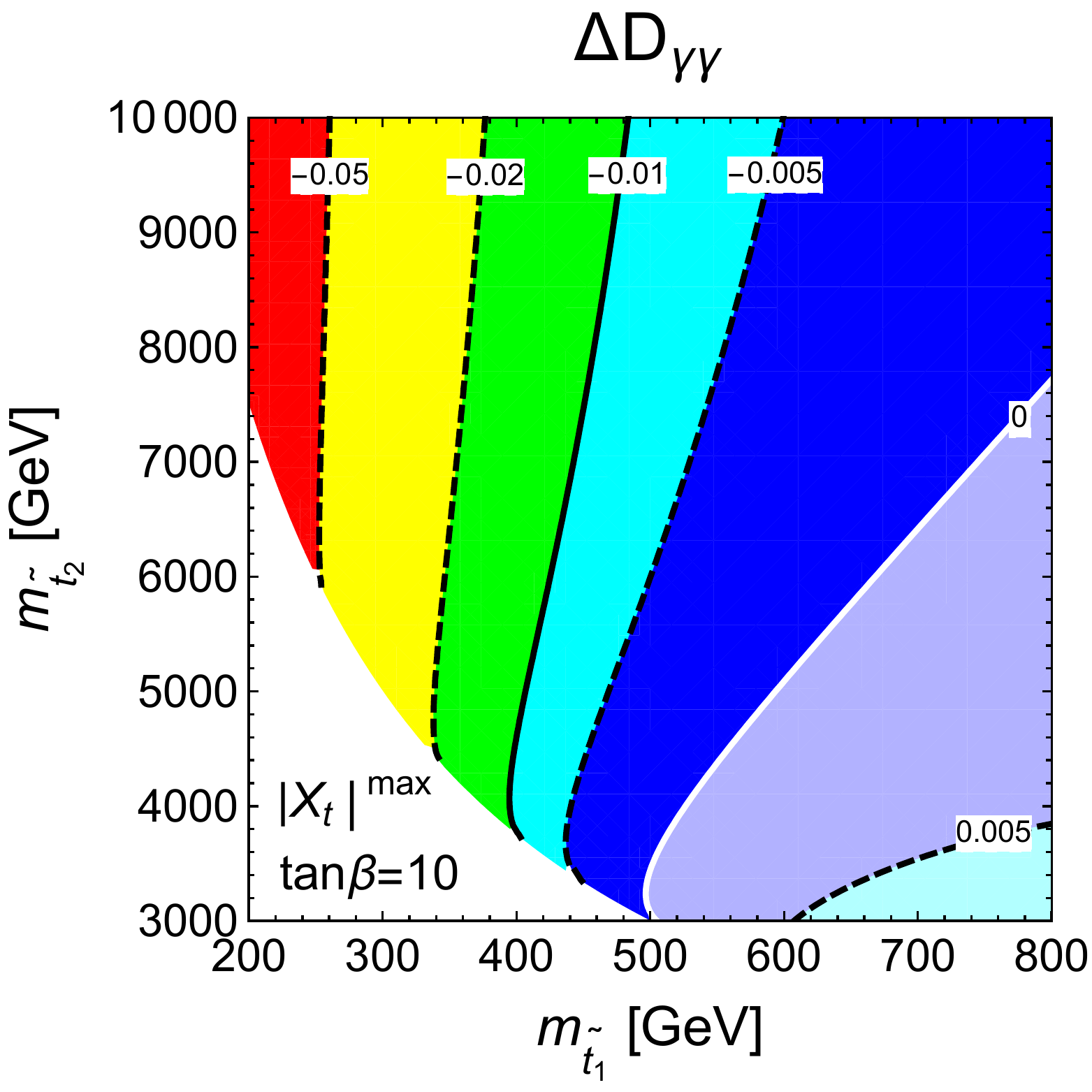} &
\includegraphics[scale=0.5]{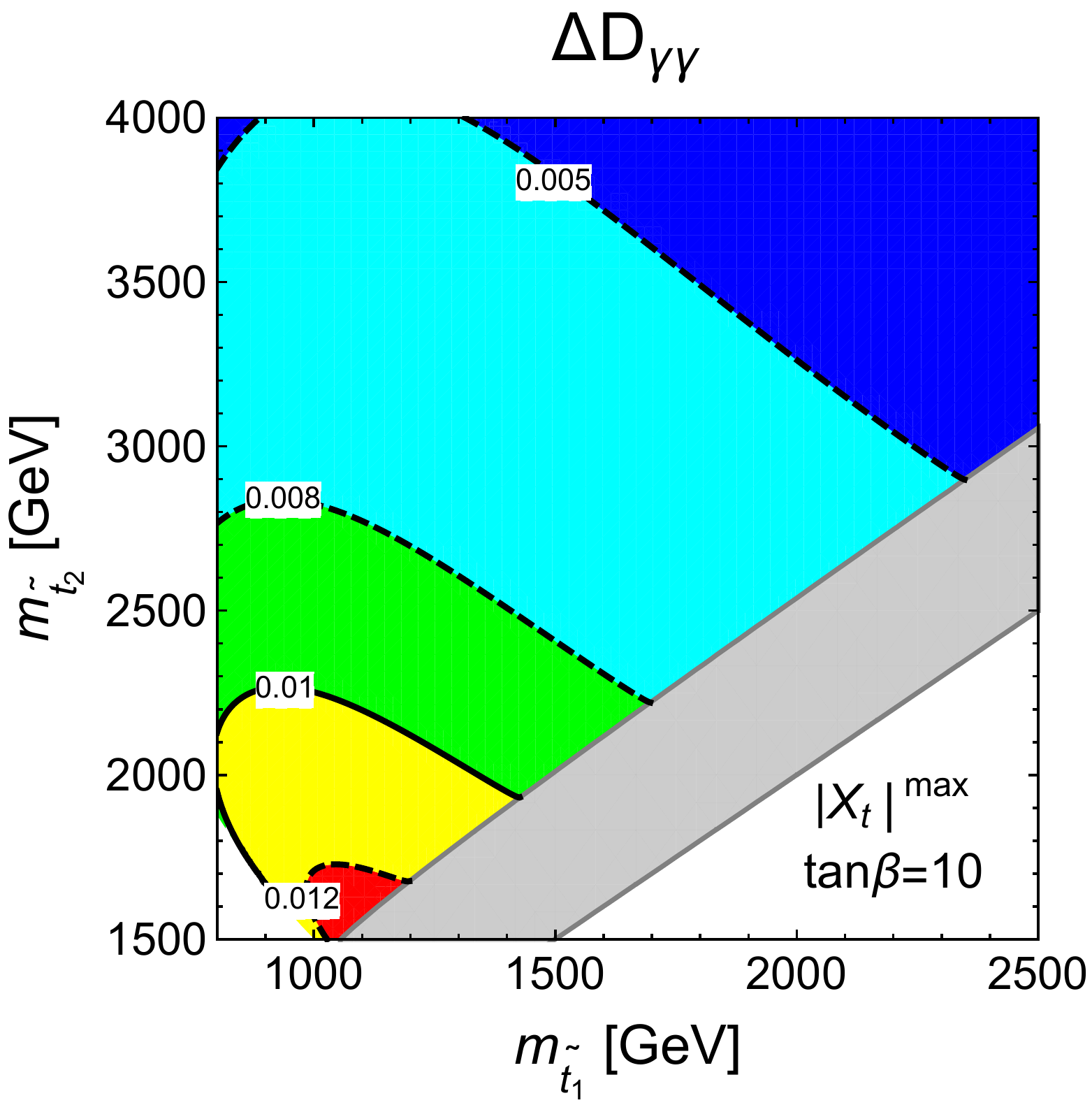}
\end{tabular}
\vspace*{-.6cm}
\caption{
Contours for the stop contributions to $D_{\gamma\gamma}$ in the $[{m_{\tilde t_1,} m_{\tilde t_2}}$] plane for the maximal possible values of
the mixing $X_t$.~The plot on the left corresponds to a large mass splitting between the two stops and the one on the right corresponds to a moderate mass splitting.}
\label{fig:stop}
\end{center}
\vspace*{-.7cm}
\end{figure}

The contributions to $D_{\gamma\gamma}$ are shown in Fig.~\ref{fig:stop} in 
the plane $[m_{\tilde t_1}, m_{\tilde t_2}]$ for the value $\tb=10$ while $X_{t}^{2}$ is fixed by the requirement that $M_h=125$ GeV when only the stop dominant contributions to the radiative corrections in the MSSM Higgs sector are considered~\cite{Anatomy}.~In this case the shift of the Higgs mass is given by,
\beq
\Delta M_h^2|^{t/\tilde{t}}_{\rm 1loop} \sim 3m_t^4/(2\pi^2v^2) [ 
\log(M_{S}^{2}/m_{t}^{2}) +X_{t}^{2}/M_{S}^{2} -X_{t}^{4}/ (12 M_{S}^{4}) ] ,
\eeq
which gives at maximum two solutions for $X_t^{2}$.~For a precision of $\Delta D_{\gamma\gamma} \approx 1\%$ one could probe stop mass values up to $m_{\tilde t_1} \sim 0.5$ TeV 
and $m_{\tilde t_2} \sim 3$ TeV for very large mass splitting between the two stops and 
$[m_{\tilde t_1}, m_{\tilde t_2}] \sim [1.5, 2]$ TeV for smaller mass splitting, when considering only the optimistic maximal solution for $X_{t}^{2}$.~These stop mass values are significantly higher than those which can be probed in direct stop pair production at the LHC, especially if the LSP neutralino is rather heavy \cite{HL-LHC}.  

\subsection*{5. Conclusions}

In this letter, we have re-emphasized the fact that the decay ratio $D_{\gamma \gamma}$ of Higgs signal rates into two photons and four charged leptons is free of theoretical uncertainties which limit the precision in other Higgs observables, but which cancel in the ratio.~ The measurement of $D_{\gamma \gamma}$ would be then limited simply by the statistical and systematic errors, which can in principle be reduced to the level of one percent at a high--luminosity LHC.~This allows us to use this Higgs decay ratio as a probe of new physics effects which is complementary to direct searches at the LHC and also other high precision electroweak observables.

We have discussed various examples, including    
anomalous Higgs couplings to top quarks and vector bosons which can be constrained at the percent level, as well as composite Higgs models where we find that compositeness scales as high as 2--3~TeV can be probed.~We have also shown that new Higgs or supersymmetric particles contributing to the H$\gamma\gamma$ loop can be probed up to  masses of several TeV
as is the case of top squarks for instance.~These scales are comparable and in some cases even higher than those accessible directly at the LHC, making this golden ratio a powerful tool 
to probe the multi--TeV scale.\bigskip 

\noindent {\bf Acknowledgements:}
We thank Adam Falkowski and Javi Serra for discussions.~AD thanks the CERN Theory Unit for its hospitality.~This work is supported by the ERC Advanced Grant {\it Higgs@LHC}.~The work of JQ was supported by the STFC Grant ST/L000326/1. 

\begin{small}

\end{small}


\begin{thebibliography}{999} 

\bibitem{NP} See for instance the summary talk of J. Ellis at the 27th International 
Symposium on Lepton Photon Interactions at High Energies, Ljubljana, Slovenia, August 2015. 
\vspace{-2mm} 

\bibitem{LHCXS} S. Dittmaier et al., the LHC Higgs cross section working group, arXiv:1101.0593.\vspace{-2mm} 

\bibitem{Baglio} J. Baglio and A. Djouadi, JHEP 1103 (2011) 055.\vspace{-2mm}  

\bibitem{HL-LHC} ATLAS collaboration, 
 arXiv:1307.7292; CMS collaboration,  
arXiv:1307.7135.\vspace{-2mm}   

\bibitem{ATLAS-all} ATLAS collaboration, Phys. Lett. B726 (2013) 88; arXiv:1507.04548.\vspace{-2mm}  

\bibitem{CMS-all} CMS collaboration,  Eur.  Phys.  J.  C75  (2015)  212.\vspace{-2mm}  

\bibitem{ATLAS-CMS} ATLAS and CMS collaborations, ATLAS-CONF-2015-044.\vspace{-2mm} 






\bibitem{ratio-old} D. Zeppenfeld, R. Kinnunen, A. Nikitenko and E. Richter-Was, 
Phys. Rev. D62 (2000) 013009; hep-ph/0002258;  
M. D\"uhrssen et al., Phys. Rev. D70 (2004) 113009.\vspace{-2mm} 

\bibitem{ratio} A. Djouadi, Eur. Phys. J. C73 (2013) 2498.\vspace{-2mm} 

\bibitem{HHG} J. Gunion, H. Haber, G. Kane and S. Dawson, ``The Higgs Hunter's
Guide", Reading 1990.\vspace{-2mm} 

\bibitem{Anatomy} A.~Djouadi,  Phys. Rept. 457 (2008) 1;
Phys. Rept. 459 (2008) 1.\vspace{-2mm} 

\bibitem{PDG} K. Olive et al., Particle Data Group, Chin. Phys. C38 (2014) 090001.\vspace{-2mm}  


\bibitem{N3LO} C. Anastasiou, Phys. Rev. Lett. 114 (2015) 21, 212001.\vspace{-2mm} 

\bibitem{ggH-NLO} M.  Spira et al.,   Nucl. Phys.  B453 (1995) 17.\vspace{-2mm} 

\bibitem{VBF-cuts} M.~Cacciari et al.,  arXiv:1506.02660 [hep-ph].\vspace{-2mm} 

\bibitem{BRerrors} \mbox{A.$\,$Denner, S.$\,$Heinemeyer, I.$\,$Puljak, D.$\,$Rebuzzi 
and M.$\,$Spira, Eur. Phys. J.C71 (2011) 1753.} For an earlier analysis see A. Djouadi, M. Spira and P. Zerwas,  Z. Phys. C70 (1996) 427.\vspace{-2mm}  

\bibitem{Fits} 
See e.g. A. Falkowski, F. Riva and A. Urbano, JHEP 1311 (2013) 111;
A. Djouadi and G. Moreau, Eur. Phys. J. C73 (2013) 2512;
J.~Ellis and T.~You, JHEP 1306 (2013) 103.\vspace{-2mm}  



\bibitem{Roberto} 
Y.~Chen, D.~Stolarski and R.~Vega-Morales, arXiv:1505.01168 [hep-ph]; 
Y. Chen, R. Harnik and R. Vega-Morales, arXiv:1503.05855 and
Phys. Rev. Lett. 113 (2014) 191801; 
Y. Chen, A. Falkowski, I. Low and R. Vega-Morales, Phys. Rev. D90 (2014) 113006;
F. Bishara et al., 
JHEP 1404 (2014) 084.\vspace{-2mm} 


\bibitem{McKeen} D. McKeen, M. Pospelov, A. Ritz, Phys.Rev. D86 (2012) 113004;
ACME Collaboration, Science 343 (2014) 269-272.\vspace{-2mm} 



\bibitem{MCHM} K. Agashe, R. Contino and A. Pomarol, Nucl. Phys. B719 (2005) 165; 
B. Gripaios et al., 
JHEP 0904 (2009) 070;
J. Mrazek et al., 
Nucl. Phys. B853 (2011) 1.\vspace{-2mm} 

\bibitem{comphiggs} 
B. Bellazzini, C. Cs\'{a}ki and J. Serra, Eur. Phys. J. C74 (2014) 2766;
M. Montull and F. Riva, JHEP 1211 (2012) 018;
J. Serra, arXiv:1506.05110 [hep-ph].\vspace{-2mm} 

\bibitem{dilaton} B. Bellazzini, C. Cs\'{a}ki, J. Hubisz, J. Serra and J. Terning, Eur.
Phys. J. C73 (2013) 2333.\vspace{-2mm} 


%
\bibitem{hMSSM} A. Djouadi, L. Maiani, G. Moreau, A. Polosa, J. Quevillon and 
V. Riquer, Eur. Phys. J. C73 (2013) 2650; JHEP 1506 (2015) 168.\vspace{-2mm}  

%
\bibitem{gamma} A. Djouadi, V. Driesen, W. Hollik and J.  Illana, 
Eur. Phys. J. C1 (1998) 149.\vspace{-2mm} 

%
\bibitem{charged} For a review on 2HDMs, see G. Branco,  Phys. Rept. 516 (2012) 1.\vspace{-2mm}  

%
\bibitem{H++} See e.g. A.~G.~Akeroyd and S.~Moretti, Phys. Rev. D86 (2012) 035015; 
A.~Arhrib, R.~Benbrik and N.~Gaur, Phys. Rev. D85 (2012) 095021; 
A.~Arhrib, R.~Benbrik, M.~Chabab, G.~Moultaka and L.~Rahili, JHEP 1204 (2012) 136.\vspace{-2mm}  

%
\bibitem{inos} J.A. Casas, J.M. Moreno, K. Rolbiecki and B. Zaldivar, JHEP 1309 (2013) 099;
B.~Batell, S.~Jung and C.~Wagner,  JHEP {\bf 1312} (2013) 075;
J. Cao et al., JHEP 1309 (2013) 043.\vspace{-2mm} 

%
\bibitem{Tau} See eg., M.~Carena, S.~Gori, N.~R.~Shah and C.~E.~M.~Wagner, JHEP
1203 (2012) 014; JHEP 1207 (2012) 175; G. Giudice, P. Paradisi and A. Strumia, JHEP 1210 
(2012) 186; U. Haisch and F. Mahmoudi, arXiv:1210.7806.\vspace{-2mm} 

%
\bibitem{Stop}  
A.~Djouadi, Phys. Lett. B435 (1998) 101; 
A. Arvanitaki, G. Villadoro, JHEP 02 (2012) 144;
A. Delgado et al.,  Eur. Phys. J. C73 (2013) 2370;
A.~Drozd et al., JHEP 1506 (2015) 028.\vspace{-2mm}  

\bibitem{pMSSM} A. Djouadi et al., hep-ph/9901246;
A. Djouadi, JL.\;Kneur and G.\;Moultaka, Comput. Phys. Commun.
176 (2007)  426; M.Muhlleitner et al.,  Comput. Phys. Commun. 168 (2005) 46.\vspace{-2mm}  

\bibitem{hdecay}  A.~Djouadi, J.~Kalinowski and M.~Spira, Comput. Phys.
Commun. 108 (1998) 56; A.Djouadi, M.Muhlleitner and M.Spira, Acta. Phys. Polon. 
B38 (2007) 635.\vspace{-2mm}  


\bibitem{Arbey} See e.g. A. Arbey et al., Phys. Lett. B708 (2012) 162; JHEP 1209 (2012) 107;
H. Baer, V. Barger and A. Mustafayev, Phys. Rev. D85 (2012) 075010; 
P. Draper et al.,  Phys. Rev. D85 (2012) 095007;
L.\;Hall, D.\;Pinner and J.\;Ruderman, JHEP 04 (2012) 131; 
S. Heinemeyer, O. Stal and G. Weiglein, Phys.Lett. B710 (2012) 201;
O.~Buchmueller et al., Eur.Phys.J. C72 (2012) 2020;
S. Akula, P. Nath and G. Peim, Phys. Lett. B717 (2012) 188;
A. Delgado, M. Garcia and M. Quiros, Phys. Rev. D90 (2014)  015016; 
J. Cao etal., JHEP10 (2012) 079. 

\end{thebibliography}
\end{document}